\documentclass[twocolumn,aps,prl,superscriptaddress,showpacs]{revtex4} 

\usepackage[english]{babel}
\usepackage[utf8]{inputenc}

\usepackage{amsmath}
\usepackage{graphicx}

\newcommand{\ie}{i.\,e.\ }

\newcommand{\eg}{e.\,g.\ }
\newcommand{\cf}{cf.\ }

\newcommand{\dd}{\;\mathrm{d}}
\newcommand{\PV}{\mathcal{P}}
\newcommand{\erf}{\mathrm{erf}}

\newcommand{\ket}[1]{\left|{#1}\right\rangle}
\newcommand{\bra}[1]{\left\langle{#1}\right|}

\newcommand{\abs}[1]{\lvert#1\rvert}           

\newcommand{\kB}{k_\mathrm{B}}

\begin{document}
\title{Attractive Optical Forces from Blackbody Radiation}
\author{M. Sonnleitner}
\affiliation{Institute for Theoretical Physics, University of Innsbruck, Technikerstra\ss e 25, A-6020 Innsbruck, Austria} 
\affiliation{Division of Biomedical Physics, Innsbruck Medical University, M\"{u}llerstra\ss e 44, A-6020 Innsbruck, Austria}
\author{M. Ritsch-Marte}
\affiliation{Division of Biomedical Physics, Innsbruck Medical University, M\"{u}llerstra\ss e 44, A-6020 Innsbruck, Austria} 
\author{H. Ritsch}
\affiliation{Institute for Theoretical Physics, University of Innsbruck, Technikerstra\ss e 25, A-6020 Innsbruck, Austria}
\begin{abstract}
	Blackbody radiation around hot objects induces ac~Stark shifts of the energy levels of nearby atoms and molecules. These shifts are roughly proportional to the fourth power of the temperature and induce a force decaying with the third power of the distance from the object. We explicitly calculate the resulting attractive blackbody optical dipole force for ground state hydrogen atoms. Surprisingly, this force can surpass the repulsive radiation pressure and actually pull the atoms against the radiation energy flow towards the surface with a force stronger than gravity. We exemplify the dominance of the ``blackbody force'' over gravity for hydrogen in a cloud of hot dust particles. This overlooked force appears relevant in various astrophysical scenarios, in particular, since analogous results hold for a wide class of other broadband radiation sources.
\end{abstract}

%
%
\maketitle
%
%
Light forces on particles microscopically arise from the basic physics of absorption and redistribution of photon momentum. For light far detuned from any optical resonance, the interaction is dominated by coherent scattering and can be attributed to an optical potential corresponding to the dynamic Stark shift of the involved atomic energy levels. Red-detuned light induces a negative Stark shift on low energy states so that particles are high-field seekers drawn towards regions of higher radiation intensity. From precision experiments in atomic spectroscopy it has been known for at least half a century that blackbody radiation also induces Stark shifts of atomic states. In particular the ground state is shifted towards lower energy~\cite{gallagher1979interactions,farley1981accurate,jentschura2008reexamining,safronova2010black,itano1982shift,angstmann2006frequency,porsev2006multipolar,mitroy2009blackbody}. Albeit a small shift, it constitutes an important perturbation of precision spectroscopy proportional to the radiated blackbody intensity~\cite{parthey2011improved} growing with the fourth power of temperature. Obviously, for a thermal source of finite size the radiation field intensity decays with distance from the surface inducing a spatially varying Stark shift. Our central claim now is that this shift constitutes a spatially varying optical potential exerting an effective optical dipole force on neutral atoms.\par
\begin{figure}
	\centering
	\includegraphics[width=8cm]{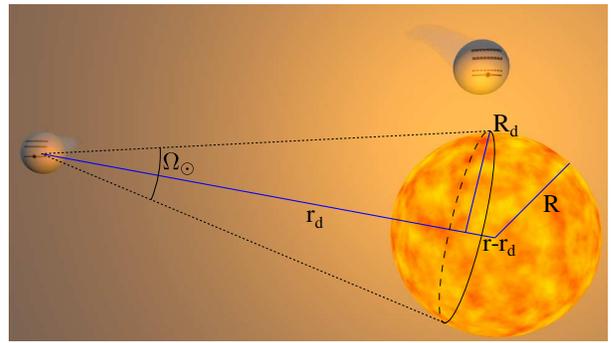}
	\caption{An artist's view of the interaction between an atom and a hot sphere of radius~$R$. From a distance~$r$ it appears as a disk of effective radius~$R_\mathrm{d}$ at distance~$r_\mathrm{d}$.}
	\label{fig:_Skizze}
\end{figure}

In this Letter we study the surprising and peculiar properties of this---so far overlooked---optical force for the simple but generic example of individual hydrogen atoms interacting with ``hot'' spheres. Since the first electronic excitation of hydrogen is in the far UV region, the largest part of a typical blackbody radiation spectrum is below the first optically coupled atomic excited state, \ie, the $2p$~state. Thus it induces an attractive dispersive dipole force analogous to the dominant force in optical tweezers~\cite{ashkin1986observation}. Of course, at the same time some narrow high frequency components are resonantly absorbed and spontaneously reemitted generating a repulsive radiation pressure. The relative size of these two components depends on temperature. As shown below, for hydrogen these forces can be explicitly calculated by generalizing known derivations of the Stark shift~\cite{farley1981accurate,jentschura2008reexamining} exhibiting a surprising dominance of the attractive dipole force up to a limiting temperature. Geometrical con\-si\-der\-ations as shown in~Fig.~\ref{fig:_Skizze} reveal that in the far field the blackbody intensity decays with the second inverse power of the distance~$r$ of the atom implying an unusual $1/r^3$ effective attractive force.\par
Let us mention here that for an atom very close to a (hot) surface additional interactions of similar magnitude such as van der Waals forces or forces arising from zero-point and thermal fluctuations of the electromagnetic field appear~\cite{antezza2005new,obrecht2007measurement}. These are also mostly attractive and depend on the detailed material properties of the surface. We will not consider these in our generic calculations.\par
In the following we will briefly review the calculation of the temperature-dependent Stark shift in a thermal field and then generalize it to the case of radiation emitted from a finite-size spherical blackbody. The corresponding potential and forces are then evaluated as a function of the radius~$R$ and the temperature~$T$ of the sphere. The resulting forces are then compared to other forces (possibly) acting on the particle, such as gravity or attraction by a dc~Stark shift from a charged sphere. We finally apply the model to atomic hydrogen moving close to a cloud of small thermal particles, where the ``blackbody optical force'' turns out to surpass the effect of gravity.\par
%
%
Let us recall some important results on polarizability and the Stark shift of light atoms and, in particular, hydrogen. The static polarizability $\alpha_{H} = (9/2) 4 \pi \varepsilon_0 a_0^3$ of ground state hydrogen atoms was calculated almost a century ago~\cite{epstein1916theorie,buckingham1937quantum}, with $a_0$ being the Bohr radius. As atomic hydrogen has its first radiative transition at an energy of $E_{2p}-E_{1s} \approx 10.2 \,\mathrm{eV}$, most of the blackbody radiation components are well below this frequency (up to temperatures of a few thousand kelvin). Hence, using this static polarizability the ground state energy shift can roughly be estimated to be $\Delta E = -\alpha_H \langle\mathbf{E}^2\rangle/2$, with the time-averaged square of the total electric field~$\mathbf{E}$; \cf Eq.~\eqref{eq:_DeltaE_H_Jentschura}. A quantitatively more reliable calculation requires summation of the perturbative contributions of all higher-lying states including their energies and dipole matrix elements, which has already been carried out by several authors~\cite{farley1981accurate,jentschura2008reexamining}.\par
The perturbative expression of the dynamic Stark shift of an energy level $\ket{n}$ in an isotropic thermal bath explicitly reads~\cite{farley1981accurate,jentschura2008reexamining}
\begin{equation}\label{eq:_DeltaE_Atom_Thermalbath}
	\Delta E_n = \frac{e^2 (\kB\, T)^3}{6 \pi^2 \varepsilon_0 (c \, \hbar)^3} \sum_{i;\, m\neq n}
	f\left(\frac{\hbar (\omega_n-\omega_m)}{\kB\, T}\right) \abs{\bra{n} r_i \ket{m}}^2,
\end{equation}
where $\hbar \omega_n$ denotes the energy of the unperturbed state $\ket{n}$, $r_i$ is the electron-core distance operator, and $f(y)$ is Cauchy's principal value integral
\begin{equation}\label{eq:_G_Integral}
	f(y)=\PV \int_0^\infty \frac{x^3}{e^x-1}\left(\frac{1}{y+x}+\frac{1}{y-x}\right) \dd\,x.
\end{equation}
In general, the sign and magnitude of the Stark shift depend on the chosen atomic or molecular state and on the radiation field temperature~$T$~\cite{farley1981accurate,jentschura2008reexamining}. However, it will mostly cause a negative energy shift for the ground state. For the $1s$ state of hydrogen it can be approximated by~\cite{jentschura2008reexamining}
\begin{equation}\label{eq:_DeltaE_H_Jentschura}
	\Delta E_{1s}(T) \approx -\frac{3 \pi^3 (\kB T)^4}{5 \alpha^3 (m_\mathrm{e} c^2)^3},
\end{equation}
which gives a small shift of $\Delta E_{1s}/\hbar \approx - 1 \,\mathrm{Hz}$ for $T=400\,\mathrm{K}$. For higher excited atomic states the blackbody shift is much larger, reaching a few $\mathrm{kHz}$~\cite{farley1981accurate}. For these states, however, the resonant absorption and emission processes discussed below also become important.\par
One arrives at similar expressions for radiation-induced absorption and stimulated transition rates between different levels resulting in an effective line width given by~\cite{farley1981accurate,jentschura2008reexamining}
\begin{equation}\label{eq:_absorption_coeff}
	\Gamma_n = \frac{e^2}{3 \pi\, c^3 \hbar \varepsilon_0} \sum_{i; m\neq n} \abs{\bra{n} r_i \ket{m}}^2 
				\frac{\abs{\omega_n-\omega_m}^3}{ \exp\left(\frac{\hbar \abs{\omega_n-\omega_m}}{\kB\, T}\right)-1}.
\end{equation}
This gives a good estimate for the expected radiation pressure force. For hydrogen in its $1s$ ground state, the temperature-dependent transition time to the $2p$ state given via $\tau_{1s \rightarrow 2p}(T) = 1/W_{1s \rightarrow 2p}(T)$ and $\Gamma_n=\sum_{m\neq n} W_{n\rightarrow m}$ is extremely long, yielding $\tau_{1s \rightarrow 2p}(300\, \mathrm{K}) \approx 10^{162}\,\mathrm{s}$ and $\tau_{1s \rightarrow 2p}(1000\, \mathrm{K}) \approx 10^{42} \,\mathrm{s}$. Note, however, that one gets a very rapid increase of this absorption with temperature, over several tens of orders of magnitude, giving $\tau_{1s \rightarrow 2p}(6000\, \mathrm{K}) \approx 0.2 \,\mathrm{s}$ around the solar temperature, where the force thus changes from attraction to repulsion.\par
For the sake of completeness let us also include the hyperfine transition within the hydrogen ground state manifold, \ie, the famous 21--cm line. For this radio transition we use the Rayleigh-Jeans approximation for the thermal energy density and the known relations between Einstein's absorption and emission coefficients to obtain the absorption rate $ W_{F0\rightarrow F1}(T) = 3 {\hbar\, \omega_{21} A_{21}/ (k_B T)}$ with $\omega_{21}\approx 8.9 \,\mathrm{GHz}$ being the angular frequency of the transition and $A_{21} \approx 2.87\times 10^{-15} \,\mathrm{s}^{-1}$~\cite{Lequeux}. We see that these rates grow linearly in temperature but are typically very small, \ie, $W_{F0\rightarrow F1}(100\,\mathrm{K})\approx 1.26\times 10^{-11} \,\mathrm{s}^{-1}$. Hence, transitions take several thousand years and only deposit negligible momentum. This shows that re\-son\-ant excitations due to thermal radiation can be neglected for neutral hydrogen in the ground state, at least up to temperatures of a few thousand $\mathrm{K}$. In this regime the force induced by the ac~Stark shift will dominate.\par
\begin{figure}
	\centering
	\includegraphics[width=8cm]{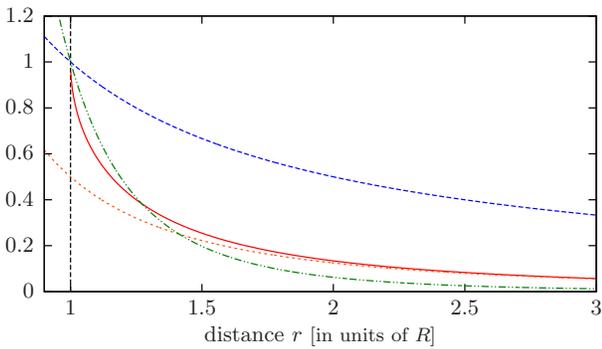}
	\caption{Comparison of the spatial decay of various interaction strengths between an atom and a hot sphere of radius~$R$: blackbody ($1-\sqrt{r^2-R^2}/r$, red solid line), gravity ($R/r$, blue dashed line) and electrostatic potential ($R^4/r^4$, green dash-dotted line), \cf also Eq.~\eqref{eq:_total_potential}. The orange dashed line extrapolates the asymptotic inverse quadratic long-range radiation potential, $R^2/(2 r^2)$.}
	\label{fig:_Vorfaktor_Potential}
\end{figure}

%
%
In an isotropic thermal bath, due to symmetry, the blackbody field cannot have any directional effect on the movement of the atom, but can only induce friction and diffusion. Thermal light radiated from a \emph{finite} source, however, creates a spatially varying Stark potential giving rise to a net force. As a generic example we consider a hot sphere of radius~$R$ with an atom located at a distance~$r\geq R$ from its center. As depicted in Fig.~\ref{fig:_Skizze} (\cf also~\cite{guess1962poynting}) the atom then sees radiation from the projection of the sphere as a disk of radius~$R_\mathrm{d}$ at distance~$r_\mathrm{d}$, with $R_\mathrm{d}=R \sqrt{1-(R/r)^2}$ and $r_\mathrm{d}=(r^2-R^2)/r $
covering a solid angle of
\begin{equation}
	\Omega_\odot = 2 \pi \left(1-\frac{r_\mathrm{d}}{\sqrt{r_\mathrm{d}^2+R_\mathrm{d}^2}}\right) = 2 \pi \left(1-\frac{\sqrt{r^2-R^2}}{r}\right).
\end{equation}
After a somewhat lengthy calculation to integrate over all incoherent contributions of the electric field radiated from a hot sphere, we finally arrive at a simple and intuitive result for the spatial dependence of the induced Stark shift,
\begin{equation}\label{eq:_DeltaE_Atom_Hotsphere_vs_Thermalbath}
	\Delta E_n^\odot (r) = \frac{1}{2}\left(1-\frac{\sqrt{r^2-R^2}}{r}\right) \Delta E_n
			= \frac{\Omega_\odot}{4\pi} \Delta E_n,
\end{equation}
where $\Delta E_n$ is the isotropic shift for a state $\ket{n}$ computed from Eq.~\eqref{eq:_DeltaE_Atom_Thermalbath}. The increasing ac~Stark shift for an atom approaching a hot sphere thus induces the rapidly growing dipole force
\begin{equation}\label{eq:_Force_Atom_Hotsphere}
	F_n^\odot(r) = - \partial_r \Delta E_n^\odot (r) = \frac{\Delta E_n}{2} \frac{R^2}{r^2 \sqrt{r^2-R^2}}.
\end{equation}
For negative $\Delta E_n$, as, \eg, for an atom in the ground state, the atom is pulled towards the sphere. Directly at the surface, where the validity of our model ends, the force diverges, but the potential energy remains finite: $\Delta E_n^\odot (R) = \Delta E_n/2$. This shift corresponds to illumination by half of the full $4\pi$ solid angle. The spatial dependence of the different potentials in Eq.~\eqref{eq:_total_potential} is shown in Fig.~\ref{fig:_Vorfaktor_Potential} with the blackbody potential falling off as $\sim R^2/(2 r^2)$ for $r\gg R$.\par
In the derivation above we set the ambient temperature around the sphere and the atom to zero. If $T_\mathrm{amb}\neq0$, the radiation from the sphere coming in from a solid angle $\Omega_\odot$ will add up with background radiation from $4\pi-\Omega_\odot$. Equation~\eqref{eq:_Force_Atom_Hotsphere} will thus change to $F_n^\odot(r) = (\Delta E_n(T)-\Delta E_n(T_\mathrm{amb})) R^2/(2 r^2 \sqrt{r^2-R^2})$. The effects of a surrounding temperature bath are thus of the order of $T^4-T_\mathrm{amb}^4$ and will be ignored in the upcoming simple examples.\par
This ``new'' attractive force is rather unexpected and---so far at least in principle---quite intriguing. In order to assess the practical importance, however, as a next step we will compare the forces created by the blackbody Stark shift to other atom-sphere interactions, such as gravitational forces or forces induced by electrostatic shifts. 
Gravitational forces may be derived from the potential
\begin{equation}\label{eq:_grav_potential}
	V_\mathrm{G}(r) = - G \frac{m\, M_\odot}{r} = - G \frac{4 \, \pi \, m \, \rho\, R^3}{3\, r} = - \frac{a_\mathrm{G} R}{r},
\end{equation}
with $M_\odot$ being the mass of the sphere of density $\rho$ and $m$ the mass of the atom, which for atomic hydrogen is $m \approx m_\mathrm{proton}$.\par
If the central sphere carries a surface charge of density~$\sigma_Q$, defined as $4 \pi R^2 \sigma_Q=Q$, the atom will experience an additional electrostatic Stark shift, which for the ground state of atomic hydrogen is found to be~\cite{Bransden}
\begin{equation}\label{eq:_level_shift_dc_Stark}
	\Delta E_{1s}^\mathrm{[Q]} = - \frac{9 a_0^3 Q^2}{16 \pi\, \varepsilon_0 r^4} = - \frac{9 \pi a_0^3 \sigma_Q^2 R^4}{\varepsilon_0 r^4} = - \frac{a_\mathrm{Q} R^4}{r^4}.
\end{equation}
In total, a neutral atom interacting with a hot sphere of radius~$R$, mass density~$\rho$, temperature~$T$, and surface charge density~$\sigma_Q$ thus sees the total potential
\begin{equation}\label{eq:_total_potential}
	V(r) = -\frac{a_\mathrm{G} R}{r} - \frac{a_\mathrm{Q} R^4}{r^4} - a_\mathrm{BB} \left(1- \frac{\sqrt{r^2- R^2}}{r} \right),
\end{equation}
where we have set $a_\mathrm{BB} = \abs{\Delta E_n}/2$. If the energy shift is positive, we must change the sign to get a repulsive interaction.\par
\begin{figure}
	\centering
	\includegraphics[width=8cm]{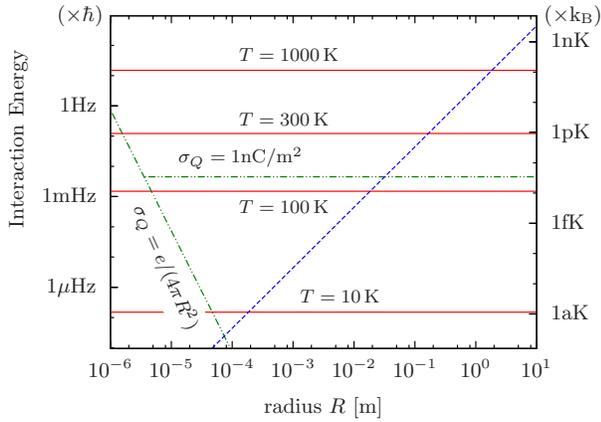}
	\caption{Comparison of the energy shift of the hydrogen ground state induced by blackbody radiation ($a_\mathrm{BB}$, red lines), the electrostatic interaction ($a_\mathrm{Q}$, green dash-dotted) and the gravitational potential energy ($a_\mathrm{G}$, blue dashed line) at the sphere surface as function of the radius~$R$ for a mass density~$\rho= 1\, \mathrm{g/cm^3}$.}
	\label{fig:_comparison_pot_energies}
\end{figure}

As forces following power laws have no natural length scale, we will use the radius~$R$ of the sphere as a reference length and first compare the potential energies generated by these different interactions at the surface. Figure~\ref{fig:_comparison_pot_energies} displays the different prefactors for different temperatures and surface charge densities as a function of sphere radius. Here, for $a_\mathrm{BB}$ we have used the approximated Eq.~\eqref{eq:_DeltaE_H_Jentschura}.\par
As the blackbody potential at the surface is independent of the sphere size, it clearly dominates gravity for small objects ($R \ll 1$m). For an interplanetary dust particle of $R=1 \,\mu\mathrm{m}$, $\rho=2 \,\mathrm{g/cm^3}$ radiating at $T = 100 \,\mathrm{K}$~\cite{Evans}, we obtain $a_\mathrm{BB}\approx 1.7\times 10^{8} a_\mathrm{G}$. For an adult human, \ie, a sphere of water with a total mass of $70 \,\mathrm{kg}$, radiating at $T = 300 \,\mathrm{K}$, $a_\mathrm{BB}\approx 0.42 \, a_\mathrm{G}$. For larger masses the blackbody potential only yields a weak perturbation; \eg, for our sun (with $R=6.96 \times 10^8 \,\mathrm{m}$, $\rho=1.408 \,\mathrm{g/cm^3}$, $T=5778 \,\mathrm{K}$~\cite{Lequeux}) we get $a_\mathrm{BB}\approx 5.5\times 10^{-15} a_\mathrm{G}$. Hence, blackbody radiation can dominate over gravity and perturb particle orbits in a way that Kepler ellipses change to rosettes or spiral trajectories towards the surface.\par
Note that aberrational effects due to the relative motion of atom and sphere generate an additional weak friction force of a different nature, leading to the known Poynting-Robertson drag on dust particles orbiting a star~\cite{guess1962poynting,Lequeux}. For fast atoms an analogous effect should be added to the gradient forces described here.\par
For the sake of brevity we also ignored the fact that thermal radiation from micron-sized particles will certainly not follow Planck's law~\cite{odashima2009mode}. A more elaborate calculation would therefore produce somewhat different numbers without changing the basic physical mechanisms and their magnitude.\par
%
%
The above considerations show that particularly strong effects can be expected from \emph{hot and light objects}. As a simple, but striking example we model a hot cloud as the integral effect of a dilute random ensemble of thermally radiating small particles. For a mass density~$\rho$ and a spherically symmetric Gaussian particle distribution of width $\sigma$, $g(r)=\exp[-r^2/(2\sigma^2)]/[(2 \pi)^{3/2} \sigma^3]$, the mean gravitational potential can be computed explicitly to give
\begin{equation}\label{eq:_random_spheres_mean_gravity}
	\langle V_\mathrm{G}(r) \rangle = - \frac{N\, a_\mathrm{G}\, R}{r} \erf\left(\frac{r}{\sqrt{2}\, \sigma}\right)
\end{equation}
with the error function $\erf(x)=2/\sqrt{\pi} \int_0^x \exp(-t^2) \dd t$. For the blackbody contribution we use the approximation $V_\mathrm{BB}(r\gg R) \simeq -a_\mathrm{BB} R^2/(2 r^2)$ to obtain
\begin{equation} \label{eq:_random_spheres_mean_BBR}
	\langle V_\mathrm{BB}(r) \rangle = -\frac{\pi\, N\, a_\mathrm{BB} \, R^2}{r}
		\PV \int_0^\infty s\, g(s) \ln \left( \frac{r+s}{\abs{r-s}} \right)\dd s.
\end{equation}
At the center of the cloud we get the simple expressions 
\begin{equation}
	\langle V_\mathrm{G}(0) \rangle = -\frac{N\, a_\mathrm{G} R}{\sqrt{\pi/2} \, \sigma}
	\quad \text{and}\quad
	\langle V_\mathrm{BB}(0) \rangle = -\frac{N\, a_\mathrm{BB} R^2}{2 \sigma^2}.
\end{equation}
The blackbody radiation induced interaction will do\-mi\-nate for parameters satisfying $\langle V_\mathrm{BB}(0) \rangle/\langle V_\mathrm{G}(0) \rangle >1$ or, making use of Eq.~\eqref{eq:_DeltaE_H_Jentschura}, when
\begin{equation}
	\frac{\sigma}{R} < \frac{\sqrt{\pi} a_\mathrm{BB}}{2\sqrt{2} a_\mathrm{G}}  
			\simeq \frac{9 \pi^{5/2} (\kB T)^4}{80 \sqrt{2} \alpha^3 G \, \rho \, (m_\mathrm{e} c^2)^3 \, m_\mathrm{p} \, R^2}.
\end{equation}
\begin{figure}
	\centering
	\includegraphics[width=8cm]{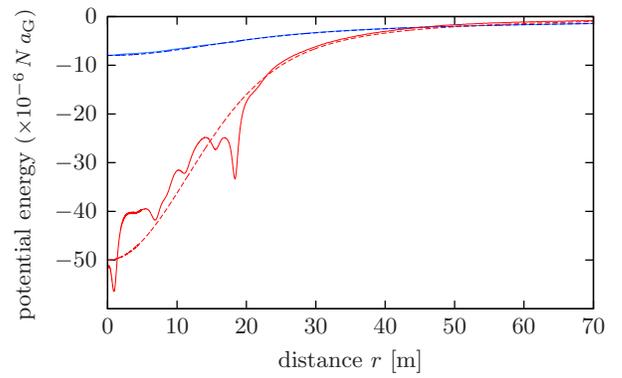}
	\caption{Gravitational [light gray (blue) lines] and blackbody induced [dark gray (red) lines] potential of randomly distributed spheres of size $R=5 \,\mu\mathrm{m}$ and temperature $T= 300\, \mathrm{K}$ such that $a_\mathrm{BB}\approx1.1\times10^9\,a_\mathrm{G}$ for $\rho=1 \,\mathrm{g/cm^3}$. The dotted lines show the potentials calculated in Eqs.~\eqref{eq:_random_spheres_mean_gravity} and~\eqref{eq:_random_spheres_mean_BBR} using a distribution of width $\sigma=300\,\mathrm{m}$. The solid lines are the result of a random sample of $N=8000$ Gaussian distributed spheres.}
	\label{fig:_random_spheres}
\end{figure}
As illustrated in Fig.~\ref{fig:_random_spheres}, we thus arrive at the quite surprising result that even for ``large'' dust clouds with $\sigma=10, \dots, 100\,\mathrm{m}$, gravitational interaction with hydrogen is not only assisted but even dominated by blackbody induced dipole forces.\par
%
%
While the idea of an attractive optical force induced by blackbody radiation appears rather exotic and unintuitive at first, we have nevertheless shown that in many cases, as, \eg, for ground state hydrogen atoms, the ``blackbody force'' dominates the repulsive radiation pressure. For small objects it can even be stronger than the gravitational interaction. Despite its outgoing radiative energy flow, a hot finite-size sphere thus attracts neutral atoms and molecules, a force, which to the best of our knowledge, has been overlooked so far. Although in many cases it will be very weak and challenging to measure in the lab, one can think of many tailored or astrophysical scenarios, which should be revisited in the context of these findings. Let us note here that at sufficiently high temperatures radiation pressure dominates and the total force changes its sign. Hence, only above a critical, rather high, temperature hydrogen atoms will be repelled by blackbody radiation as intuitively expected. Note that the dipole force is state selective and can induce spatial separation of atoms in different long-lived states.\par
Our results actually go beyond the case of blackbody radiation and basically also hold for other broadband incoherent light sources, with narrower frequency distributions but higher photon flux compared to blackbody radiation. This potentially generates much stronger forces which are strongly species specific. On the microscopic scale it is also important to note that the precise shape of the hot particle surface will be decisive and strong enhancement effects could be expected near tips, grooves, and edges. Hence, our findings could go much beyond the originally intended scope, \eg, towards the effect of hot microstructured surfaces in vacuum chambers or the total energy shift induced by cosmic background radiation.\par
%
%
\acknowledgments
We thank Sabine Schindler and Josef St\"{o}ckl for stimulating discussions, Miles Padgett and Jean Dalibard for helpful comments and acknowledge support by the ERC Advanced Grant (Project No.\ 247024 catchIT).
%
%

%
\end{document}